\documentclass{article}

\usepackage{arxiv}

\usepackage[utf8]{inputenc}
\usepackage[T1]{fontenc}
\usepackage{hyperref}
\usepackage{url}
\usepackage{booktabs}
\usepackage{amsfonts}
\usepackage{nicefrac}
\usepackage{microtype}
\usepackage{lipsum}
\usepackage{float}
\usepackage{graphicx}
\usepackage{doi}
\usepackage{amsmath}
\usepackage{amssymb}
\usepackage{xcolor}
\usepackage{subcaption}
\usepackage{tikz}
\usepackage{quantikz}
\usetikzlibrary{arrows.meta, positioning}
\definecolor{darkgreen}{RGB}{0,100,0}

\makeatletter
\renewcommand{\@maketitle}{
  \newpage
  \null
  \vskip 2em
  \begin{center}
    {\LARGE\bfseries
      \hrule height 1.2pt
      \vspace{0.8em}
      \@title
      \vspace{0.8em}
      \hrule height 1.2pt
    }
    \vskip 1.5em

    \begin{minipage}{0.98\textwidth}
      \centering
      Gal G. Shaviner$^{\text{a}}$ \quad 
      Ziv Chen$^{\text{b,c}}$ \quad 
      Steven H. Frankel$^{\text{a,c}}$ \par
      \vspace{0.5em}
      {\small
      $^{\text{a}}$Faculty of Mechanical Engineering, Technion – Israel Institute of Technology, Haifa, Israel 3200003 \par
      $^{\text{b}}$Faculty of Electrical and Computer Engineering, Technion – Israel Institute of Technology, Haifa, Israel 3200003 \par
      $^{\text{c}}$Helen Diller Quantum Center, Technion – Israel Institute of Technology, Haifa, Israel 3200003 \par
      \texttt{\{gal.shaviner, ziv.chen\}@campus.technion.ac.il}, \texttt{frankel@me.technion.ac.il}
      }
    \end{minipage}

    \vskip 1em
    {\@date}
  \end{center}
  \par
  \vskip 1.5em
}
\makeatother

\title{Quantum Singular Value Transformation for Solving the Time-Dependent Maxwell's Equations}
\author{} 
\date{}   

\begin{document}
\maketitle

\begin{abstract}
{This work presents a quantum algorithm for solving linear systems of equations of the form \(\mathbf{A}{\frac{\mathbf{\partial f}}{\mathbf{\partial x}}} = \mathbf{B}\mathbf{f}\), based on the Quantum Singular Value Transformation (QSVT). The algorithm uses block-encoding of \(A\) and applies an 21st-degree polynomial approximation to the inverse function \(f(x) = 1/x\), enabling relatively shallow quantum circuits implemented on 9 qubits, including two ancilla qubits, corresponding to a grid size of 128 points. Phase angles for the QSVT circuit were optimized classically using the Adagrad gradient-based method over 100 iterations to minimize the solution cost. This approach was simulated in PennyLane and applied to solve a 1D benchmark case of Maxwell’s equations in free space, with a Gaussian pulse as the initial condition, where the quantum-computed solution showed high fidelity of more than 99.9 \% when compared to the normalized classical solution. Results demonstrate the potential of QSVT-based linear solvers on simulators with full quantum state access. However, practical hardware implementations face challenges because accessing the complete quantum state is infeasible. This limitation restricts applicability to cases where only $O({poly}(n))$ observables are needed. These findings highlight both the promise and current limitations of using quantum algorithms, such as QSVT, to solve linear systems of equations, and they point to the need for the development of measurement-efficient algorithms for near-term quantum devices.}
\end{abstract}


\section{Introduction}
Maxwell’s equations and the Lorentz force law form the foundation of classical electromagnetism, underpinning technologies such as optics, wireless communication, and electronics \cite{lam2003wireless, makarenko2020generalized, thompson2012maxwell}. Due to their inherent nonlinearity and hyperbolic nature, obtaining exact analytical solutions to these equations is challenging, particularly in multidimensional scenarios. As a result, computational approaches such as the finite-difference time-domain (FDTD) technique \cite{JENKINSON2018192}, finite-element methods \cite{CHANG198889}, and finite-volume schemes \cite{HERMELINE20089365} have become essential tools for producing approximate numerical solutions.

These numerical methods often rely on solving large systems of linear equations, which arise naturally from the discretization of Maxwell’s equations and many other partial differential equations. Linear systems of equations play an important role in many areas of science and technology. For example, in biology they arise in modeling metabolic networks, population dynamics, and biochemical reaction pathways~\cite{J_Merrill_2019,https://doi.org/10.1111/j.1939-7445.1988.tb00046.x,ALBERTY1996507}. In machine learning, training algorithms such as linear regression or support vector machines often require solving systems of equations derived from optimization objectives~\cite{8903465}. In computational physics and engineering, discretizing partial differential equations using finite difference or finite element methods reduces the problem to solving large, sparse linear systems~\cite{GAO2014115}. In fluid dynamics, the discretization and linearization of the Navier–Stokes equations during numerical simulation also necessitates robust and efficient solvers~\cite{DAS2023105793}. The ubiquity of linear systems across scientific disciplines highlights the fundamental importance of developing scalable algorithms for their solution.

In the past decade, significant attention has been given to the possibility of solving linear systems on quantum computers. Classically solving an N $\times$ N linear system (N equations for N unknowns) scales polynomially in N. In contrast, Harrow-Hassidim-Lloyd (HHL) \cite{PhysRevLett.103.150502} introduced a quantum algorithm that scales logarithmically in N, suggesting that quantum computers may provide an exponential speedup for certain linear system problems. More recently, the Variational Quantum Linear Solver (VQLS) \cite{Bravo_Prieto_2023} has been proposed as a hybrid quantum-classical alternative that is more suitable for near-term quantum devices. VQLS formulates the solution of $Ax = b$ as a variational optimization problem, where a parameterized quantum circuit prepares an approximate solution state $|x\rangle$ by minimizing a cost function that encodes the residual error in the linear system. This method avoids deep quantum circuits and phase estimation, making it more practical under realistic noise and coherence constraints. However, VQLS relies on repeated ansatz optimization, which can be computationally expensive and less practical for solving time-dependent PDEs that require solutions at multiple time steps. An alternative, fully quantum paradigm is the Quantum Singular Value Transformation (QSVT) \cite{Gily_n_2019}, which provides a powerful and modular framework for implementing matrix functions on quantum states. QSVT leverages block-encoded matrices and polynomial approximations to perform transformations directly on the singular values of an operator. By constructing a polynomial approximation to the reciprocal function $f(x) \approx 1/x$ over the spectrum of $A$, QSVT enables the implementation of an inverse transformation $A^{-1}$ as a quantum circuit. This approach generalizes and improves upon earlier linear system solvers by avoiding phase estimation and offering provable bounds on approximation error and complexity, while remaining compatible with fault-tolerant quantum computation.

Several recent works have addressed the use of QSVT for solving partial differential equations. Lubasch et al. \cite{lubasch2025quantumcircuitspartialdifferential} leveraged QSVT alongside the quantum Fourier transform (QFT) to design quantum circuits with optimal asymptotic gate complexity. Their constructions, applied to canonical PDEs such as incompressible advection, heat, isotropic acoustic wave, and Poisson’s equations, yield circuits that are conceptually simple and hardware-efficient. Novikau et al. \cite{Novikau_2023} proposed a quantum algorithm using QSVT to simulate dissipative wave propagation in inhomogeneous linear media as a boundary-value problem. Their approach constructs quantum circuits modeling electromagnetic wave propagation in one-dimensional systems with outgoing boundary conditions.

When it comes to Maxwell's equations, several studies have investigated their solution using quantum computing approaches. Jin et al. \cite{jin2023quantumsimulationmaxwellsequations}   proposed quantum algorithms based on the Schrödingerisation technique, enabling unitary evolution for simulating Maxwell's equations. Chen et al. \cite{chen2025quantumphysicsinformedneuralnetworks} used Quantum Physics Informed Neural Networks (QPINN) to solve two-dimensional (2D)
time-dependent Maxwell’s equations. Their approach utilized a parameterized quantum circuit
(PQC) in conjunction with the classical neural network architecture while enforcing physical
laws, including a global energy conservation principle, during the training of the neural network. Nguyen and Thompson \cite{nguyen2024solvingmaxwellsequationsusing} investigated the potential of using the variational quantum imaginary time evolution (VarQITE) algorithm on near-term quantum hardware to solve for the Maxwell’s equations. Koukoutsis et al. \cite{Koukoutsis_2023} developed a quantum Schrödinger representation of Maxwell’s equations tailored for wave propagation in cold, inhomogeneous magnetized plasmas, enabling unitary, energy-preserving evolution. They also proposed a qubit lattice algorithm (QLA) that can be implemented on quantum computers and tested classically, demonstrated with simulations of electromagnetic wave scattering.

The central idea of QSVT is to encode a target matrix into a block of a unitary operator, then apply a sequence of single-qubit rotations and controlled operations to approximate a polynomial function of its singular values. This framework includes several quantum algorithms, including quantum walks, amplitude amplification, and Hamiltonian simulation, under a single algebraic technique. Unlike HHL, QSVT-based methods do not require explicit eigenvalue estimation or ancilla post-selection, which can significantly reduce circuit depth and improve practicality on near-term quantum devices.

Recent theoretical work \cite{bergholm2022pennylaneautomaticdifferentiationhybrid} shows that QSVT achieves nearly optimal query complexity for various matrix functions, including the inverse, sign, and exponential functions, depending on the spectral properties of the matrix. For the linear system problem, approximating the inverse function $f(x) = 1/x$ over the spectrum of $A$ leads to efficient condition number-dependent performance. This makes QSVT a compelling candidate for quantum linear solvers, especially when combined with improved block-encoding constructions and noise-resilient circuit compilation techniques.

This work implements a QSVT-based quantum algorithm to solve $Ax = b$ for a non-Hermitian matrix $A$. Full simulations are performed in PennyLane \cite{bergholm2022pennylaneautomaticdifferentiationhybrid}, optimizing a QSVT polynomial approximation to the inverse function and demonstrating high-fidelity recovery of the normalized classical solution.

Despite the theoretical advantages, practical implementation of QSVT faces challenges such as precision loss due to high-degree polynomials and resource overhead from controlled-unitary decompositions. In this work, we address these by focusing on small-scale simulations and adopting numerical optimization to construct low-degree polynomial approximations of the inverse. By leveraging PennyLane's hybrid quantum-classical framework, we simulate the full QSVT circuit, verify the accuracy of the encoded solution, and choose parameter regimes that preserve fidelity.

\section{Maxwell's equations formulation}
Maxwell's microscopic equations can be written as:
\begin{equation}
\nabla \cdot E= \frac{\rho}{\varepsilon}   
\end{equation}
\begin{equation}
\nabla \cdot B = 0
\end{equation}
\begin{equation}
\nabla \times E = - \frac{\partial B}{\partial t}
\end{equation}
\begin{equation}
\nabla \times B = \mu \left(J + \varepsilon\frac{\partial E}{\partial t}\right)
\end{equation}
With  
$\mathbf{E}$ the electric field,  
$\mathbf{B}$ the magnetic field,  
$\rho$ the electric charge density, and  
$\mathbf{J}$ the current density.  
$\varepsilon$ is the permittivity, and  
$\mu$ is the permeability.

Assuming no free charges or currents, the Maxwell’s equations can be written:
\begin{equation}
\nabla \times B = \mu \varepsilon\frac{\partial E}{\partial t}
\end{equation}
\begin{equation}
\nabla \times E = - \frac{\partial B}{\partial t}
\end{equation}
More generally, for linear materials, the constitutive relations between the electric field $E$, the magnetic field $B$, the displacement field $D$ and the magnetizing field $H$ are:
\begin{equation}
D = \varepsilon E \quad, \quad B = \mu H
\end{equation}
where $\varepsilon$ is the permittivity and $\mu$ the permeability of the material.
By substituting B with H, Eqs. 5-6 are transformed into the formulation of the electric field E and the magnetizing field $H$:

\begin{equation}
\frac{\partial E}{\partial t} = \frac{1}{\varepsilon} \nabla \times H
\end{equation}
\begin{equation}
\frac{\partial H}{\partial t} = - \frac{1}{\mu} \nabla \times E
\end{equation}

Maxwell's Equations in 1D:

\begin{equation}
\frac{\partial E_x}{\partial t} = - \frac{1}{\varepsilon} \frac{\partial H_y}{\partial z} 
\end{equation}
\begin{equation}
     \frac{\partial H_y}{\partial t} = - \frac{1}{\mu} \frac{\partial E_x}{\partial z}
\end{equation}
And in free space where $\varepsilon=\mu=1$ Eqs. 10-11 become:
\begin{equation}
\frac{\partial E_x}{\partial t} = - \frac{\partial H_y}{\partial z} 
\end{equation}
\begin{equation}
     \frac{\partial H_y}{\partial t} = - \frac{\partial E_x}{\partial z}
\end{equation}

\section{Methods}
One of the most common ways to represent a linear system is in matrix-vector form: given an invertible matrix \( A \in \mathbb{R}^{N \times N} \) and a vector \( \vec{b} \), the aim is to solve \( A \vec{x} = \vec{b} \), which requires computing \( \vec{x} = A^{-1} \vec{b} \). Using QSVT, the inverse operation is approximated by applying a polynomial transformation to the singular values of \( A \). In practice, this means constructing a polynomial \( P(t) \approx \frac{1}{t} \), implemented through a sequence of controlled unitaries parameterized by phase angles. This section deals with the construction of such polynomial approximations using Quantum Singular Value Transformation (QSVT), detailing the design of the phase factors and the implementation of the corresponding quantum circuits for approximating the inverse of \( A \).

\subsection{Padé scheme}
The numerical solution for Maxwell’s equations involves implementing a fourth order Padé \cite{LELE199216} scheme to calculate the spatial derivatives in the RHS of Maxwell’s equations in the following way

\begin{equation}
\beta f'_{i-2} +\alpha f'_{i-1} + f'_i + \alpha f'_{i+1} + \beta f'_{i+2} 
= 
c\frac{(f_{i+3} - f_{i-3})}{6h} 
+
b\frac{(f_{i+2} - f_{i-2})}{4h} 
+
a\frac{(f_{i+1} - f_{i-1})}{2h} 
\end{equation}

The coefficients for the classical Padé scheme are:

\[
\alpha = \frac{1}{4} \quad,\quad\beta = 0 \quad,\quad a = \frac{3}{2} \quad,\quad b=0 \quad,\quad c=0
\]

The following linear system of equations is solved to calculate the first derivative $f'$:

\begin{equation}
\frac{1}{4} f'_{i-1} + f'_i + \frac{1}{4} f'_{i+1} = \frac{3}{4h}(f_{i+1} - f_{i-1}) 
\end{equation}

Writing this discrete equation at every grid point results in the need to solve a system of linear algebraic equations to compute the discrete derivative, and it can be represented in the following form
\begin{equation}
 A \mathbf{f}' = B \mathbf{f}
\end{equation}

Once the derivative $\mathbf{f}'$ is calculated, the system can be advanced in time using Euler time stepping \cite{Griffiths2010}.

\subsection{Quantum Circuit}

\begin{figure}[H]
  \centering
  \begin{subfigure}[t]{1.0\textwidth}
    \centering
    \includegraphics[width=1.0\linewidth]{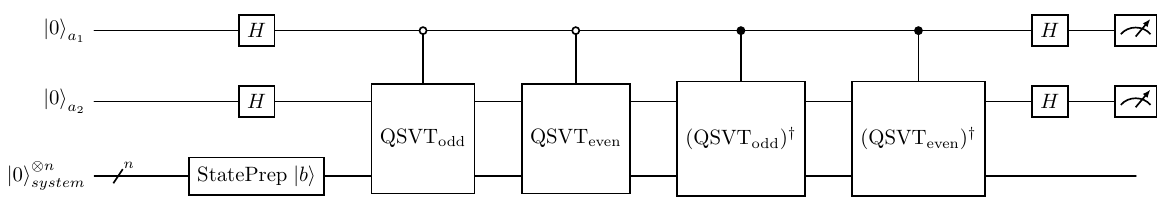}
    \caption{}
    \label{fig:higher_level_circ}
  \end{subfigure}\hfill
  
  \medskip

  \begin{subfigure}[t]{1.0\textwidth}
    \centering
    \includegraphics[width=1.0\linewidth]{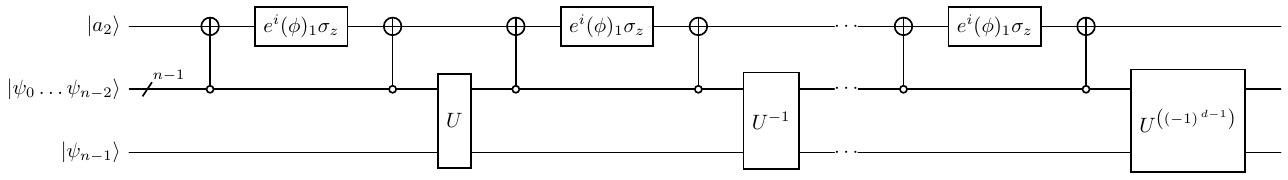}
    \caption{}
    \label{fig:qsvt_lower_circ}
  \end{subfigure}\hfill
  
  \caption{(a) Higher level view of the circuit used for the QSVT algorithm. The upper two qubit lines are the two ancilla qubits ($a_1$ and $a_2$), the \textit{StatePrep $\ket{b}$} gate encodes state \textit{b} (from $Ax=b$) as the state \ket{b} into the system qubits. A filled dot in a control gate is activated when the state is \ket{1}, and an empty dot is activated when the state is \ket{0}. The two ancillary qubits are measured and reset at the end of the circuit. (b) Decomposition of the QSVT gates, where $QSVT_{even}$ and $QSVT_{odd}$ are built according to Eqns. \ref{eqn:qsvt_even}, \ref{eqn:qsvt_odd}. Here $\sigma_z$ is the Pauli-Z matrix, the controlled not gates have $n-1$ qubits as control so they behave as "many-qubit-Toffoli-not" gates. $U$ is determined according to the block encoding for Eqn. \ref{eqn:block}.}
    \label{fig:circuit}
\end{figure}

The circuits shown in Fig. \ref{fig:circuit} implement the QSVT protocol.
In \ref{fig:higher_level_circ}, the circuit implements the whole QSVT protocol, using two ancilla qubits, where the first ancilla ($a_1$) determines whether the real or imaginary part of $U$ (Eqn. \ref{eqn:real_and_imag}) is calculated and the second one ($a_2$) determines whether the calculated degree is even or odd.
controlled sequence of unitary operations interleaved with phase rotations. The top qubit, initialized in the state $\ket{0}^{\otimes n}$, acts as a control register. Each block applies a controlled-$\sigma_z$ phase rotation $e^{i \phi_j \sigma_z}$ conditioned on the control qubit, followed by controlled operations $U$ and its inverse $U^{-1}$ acting on the lower qubits. This pattern repeats $d$ times (d is the polynomial degree), with the final controlled operation raised to the power $((-1)^{d-1})$.

\subsection{Algorithm}
The proposed algorithm solves the 1D Maxwell's equations by combining the Padé finite difference scheme with QSVT. At each time step, spatial derivatives are computed using the compact Padé approximation, written in matrix form as $A \frac{\partial f}{\partial z} = Bf$, where ${f} = {E}_x$ or ${f} = {H}_y$. The quantity $ {A^{-1}Bf} $ is computed via QSVT. The process begins by calculating \( \frac{\partial H_y}{\partial z} \), then advancing the electric field \( E_x \) in time using the explicit Euler method. Next, \( \frac{\partial E_x}{\partial z} \) is calculated, and the magnetic field \( H_y \) is advanced in time as well using the explicit Euler method. These four steps are repeated to evolve the electromagnetic fields in time, The loop terminates when the simulation time t exceeds the final time $t_{final}$, ending once $t \leq t_{final}$ is no longer true.

\begin{figure}[H]
    \centering
    \includegraphics[width=0.5\linewidth]{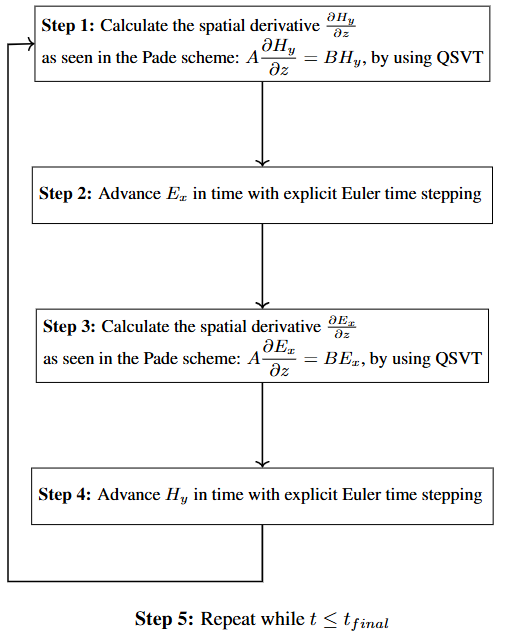}
    \caption{Block diagram representation of the proposed algorithm.}
    \label{algorithm}
\end{figure}

A Block diagram representation of the proposed algorithm is shown in Fig. \ref{algorithm}.

\subsection{Polynomial design}
To implement QSVT, a polynomial approximation of the function \( f(x) = \frac{1}{x} \) is constructed over the interval \( x \in \left[ \frac{1}{\kappa}, 1 \right] \), where \( \kappa \) denotes the condition number of the matrix \( A \). Since QSVT supports only polynomial transformations bounded within the interval \( [-1, 1] \), the approximation targets the scaled function \( s \cdot \frac{1}{x} \), with the scaling factor \( s \) chosen to ensure that the resulting polynomial remains bounded. The values $\kappa =4$ and $s=0.10145775$ were chosen as seen in \cite{bergholm2022pennylaneautomaticdifferentiationhybrid}.

To solve Maxwell's equations a 21st-order polynomial is used, resulting in a sum of even and odd polynomials of degrees 10 and 11, respectively. This order was chosen because it required relatively low computational effort while delivering high-fidelity results. The reason for using an even and odd polynomials is explained in Sec. \ref{subsec:phase-angles}.

\subsection{Phase angles}
\label{subsec:phase-angles}
The QSVT operation as used in PennyLane, is fully differentiable. This property enables determining phase angles which are needed for block encoding (explained in Sec. \ref{subsec:block_encoding}) by employing gradient-based optimization techniques. To define the QSVT transformation, a total of 21 phase angles are randomly initialized and then iteratively optimized by minimizing the loss function in Eq.~\ref{eqn:loss} using the Adagrad optimizer \cite{nanni2021} with a learning rate of 0.1.

A single QSVT circuit yields a transformation with fixed parity. To achieve a more accurate approximation, a linear combination of even and odd polynomials is used as seen in \cite{Gily_n_2019, bergholm2022pennylaneautomaticdifferentiationhybrid}. This is accomplished via a linear combination of unitaries (LCU). The phase angles are first separated into two groups corresponding to even and odd parity. One of the ancilla qubits is then prepared in an equal superposition state. Each QSVT transformation, either even or odd, is applied conditionally based on the state of the ancilla. Finally, the ancilla qubit is reset.

To enable this construction, the signal-processing operator \cite{generalizedquantumsignalprocessing} \( \Pi(\phi) \), which implementation will be explained in Sec. \ref{subsec:block_encoding}, must be generalized to higher dimensions. This is achieved by defining a diagonal unitary that applies the phase \( e^{i\phi} \) to the subspace associated with the encoded block and \( e^{-i\phi} \) to the orthogonal subspace. The corresponding signal-processing operator is referred to as projector-controlled phase gates, and is denoted by the symbol \( \Pi(\phi) \).


\begin{equation}
\Pi(\phi_k) =
\begin{bmatrix}
e^{(-1)^{b_1} i \phi_k} & 0 & 0 & \cdots & 0 \\
0 & e^{(-1)^{b_2} i \phi_k} & 0 & \cdots & 0 \\
0 & 0 & e^{(-1)^{b_3} i \phi_k} & \cdots & 0 \\
\vdots & \vdots & \vdots & \ddots & \vdots \\
0 & 0 & 0 & \cdots & e^{(-1)^{b_{2N}} i \phi_k}
\end{bmatrix}_{2N \times 2N}
\end{equation}

Where $b_i$ is 0 if the subspace is associated with the encoded block and 1 otherwise, $N=2^{n}$, the polynomial degree $d=21$ and $k=1,\dots,d$.

\subsection{Matrix and Block Encoding}
\label{subsec:block_encoding}
The first step of the process involves block-encoding \( A \) into a unitary matrix. This work used the quantum signal processing (QSP) \cite{QSP} method to implement this process. Alternate products of \( U(A) \) and \( \Pi(\phi) \), keeping \( A \) fixed and varying the phase angles \( \phi_k \), result in a matrix whose top-left corner is a polynomial transformation of \( A \). Mathematically, the QSP transformation is expressed as

\begin{equation}
\Pi(\phi_0) \prod_{k=1}^{d} U(A) \Pi(\phi_k) = 
\begin{pmatrix}
P(A) & * \\
* & *
\end{pmatrix}
\end{equation}

where the asterisks indicate matrix entries that are not of interest. The intuition behind this result is that each application of \( U(A) \) increases the polynomial degree of the transformation, and by interleaving signal-processing operators \( \Pi(\phi_k) \), the coefficients of the resulting polynomial can be tuned. However, for high-degree polynomials, the resulting circuit becomes increasingly deep, which can introduce significant noise and errors due to hardware limitations in near-term quantum devices.

Mathematically, when the polynomial degree \( d \) is even (i.e., the number of phase angles is \( d + 1 \)), the QSVT algorithm defines the transformation as
\begin{equation}
\label{eqn:qsvt_even}
P_{even}=\left[ \prod_{k=1}^{d/2} \widetilde{\Pi}_{\phi_{2k-1}} U(A)^\dagger \Pi_{\phi_{2k}} U(A) \right] \Pi_{\phi_{d+1}} = \begin{pmatrix}
P_{even}(A) & * \\
* & *
\end{pmatrix}
\end{equation}
where $\Tilde {\Pi}$ indicates projector-controlled phase gates that act differently depending on whether it is applied to the column or row subspace. 

The polynomial transformation \( P(A) \) is expressed using the singular value decomposition of \( A \) as
\begin{equation}
P(A) = \sum_k P(\sigma_k) \, |w_k\rangle \langle v_k|
\end{equation}
where \( \sigma_k \) are the singular values, and \( |v_k\rangle \), \( |w_k\rangle \) are the corresponding right and left singular vectors, respectively, written in Dirac (braket) notation.

For technical reasons, when the polynomial degree \( d \) is odd, the sequence takes a slightly different form:
\begin{equation}
\label{eqn:qsvt_odd}
P_{odd}=\widetilde{\Pi}_{\phi_1} \left[ \prod_{k=1}^{(d-1)/2} \Pi_{\phi_{2k}} U(A)^\dagger \widetilde{\Pi}_{\phi_{2k+1}} U(A) \right] \Pi_{\phi_{d+1}} = 
\begin{pmatrix}
P_{odd}(A) & * \\
* & *
\end{pmatrix}
\end{equation}

$P_{tot}(A)$ is represented as the sum of $P_{even}(A)$ and $P_{odd}(A)$:
\begin{equation}
\label{eqn:p_tot}
P_{tot}(A)=P_{even}(A)+P_{odd}(A)
\end{equation}

it is possible to express the real part of a complex number $z$ as $Re[z]=\frac{1}{2}(z+z^*)$, so in the same manner for $P_{tot}$:
\begin{equation}
\label{eqn:real_and_imag}
P_{tot,\;real}(A)=\frac{1}{2}(P_{tot}+P_{tot}^*)
\end{equation}

To ensure that U(A) is hermitian, it is defined as in~\cite{bergholm2022pennylaneautomaticdifferentiationhybrid}:

\begin{equation}
U(A) =
\begin{bmatrix}
A & \sqrt{I - A A^\dagger} \\
\sqrt{I - A^\dagger A} & -A^\dagger
\end{bmatrix}
\label{eqn:block}
\end{equation}

This and the polynomial approximation are implemented using PennyLane’s \texttt{qml.BlockEncode}, allowing the inverse to be constructed efficiently within the quantum circuit. It is important to note that there is no real implementation behind the BlockEncode; it is just a mathematical operation of matrix multiplication and therefore only suitable for a simulator. A common approach is to represent the operator \( A \) as a linear combination of unitaries and define associated \text{PREPARE} and \text{SELECT} operators; the operator \( U = \text{PREPARE}^\dagger \cdot \text{SELECT} \cdot \text{PREPARE} \) as seen in \cite{simon2025ladderoperatorblockencoding} then forms a block-encoding of \( A \) up to a constant factor.

\subsection{Classical optimization}
The loss function in Eq. \ref{eqn:loss} is defined as the mean squared error between the real part of the \((0,0)\) (Eq. \ref{eqn:block}) element of QSVT matrix output and the target function \( f(x) = \frac{s}{x} \) evaluated at a discrete set of sampled points within the domain of interest. This formulation enables determining the phase angles which are needed for block encoding.
\begin{equation}
\mathcal{L}(\phi) = \frac{1}{M} \sum_{i=1}^M \left( \operatorname{Re}\big( [P_{tot}(x_i, \phi)]_{0,0} \big) - \frac{s}{x_i} \right)^2
=
\frac{1}{M} \sum_{i=1}^M \left( \operatorname{Re}\big( P(A) \big) - \frac{s}{x_i} \right)^2
\label{eqn:loss}
\end{equation}
where:
\begin{itemize}
    \setlength\itemsep{2pt} 
    \setlength\parskip{0pt} 
    \setlength\parsep{0pt}  
    \item \(M\) is the number of sampled points, in our case \(M = 100\).
    \item \(x_i\) are the samples in the interval \(\left[\frac{1}{\kappa}, 1\right], \) where \(\,\, 1 \leq i \leq M. \) 
    \item \(P(A)\) is the polynomial evaluated at \(x_i\) with the phase angles parameters \(\phi\).
    \item \(\frac{s}{x_i}\) is the target function value.
\end{itemize}

\section{Results and analysis}

\subsection{Phase angles optimization}
\begin{figure}[H]
    \centering
    \includegraphics[width=0.5\linewidth]{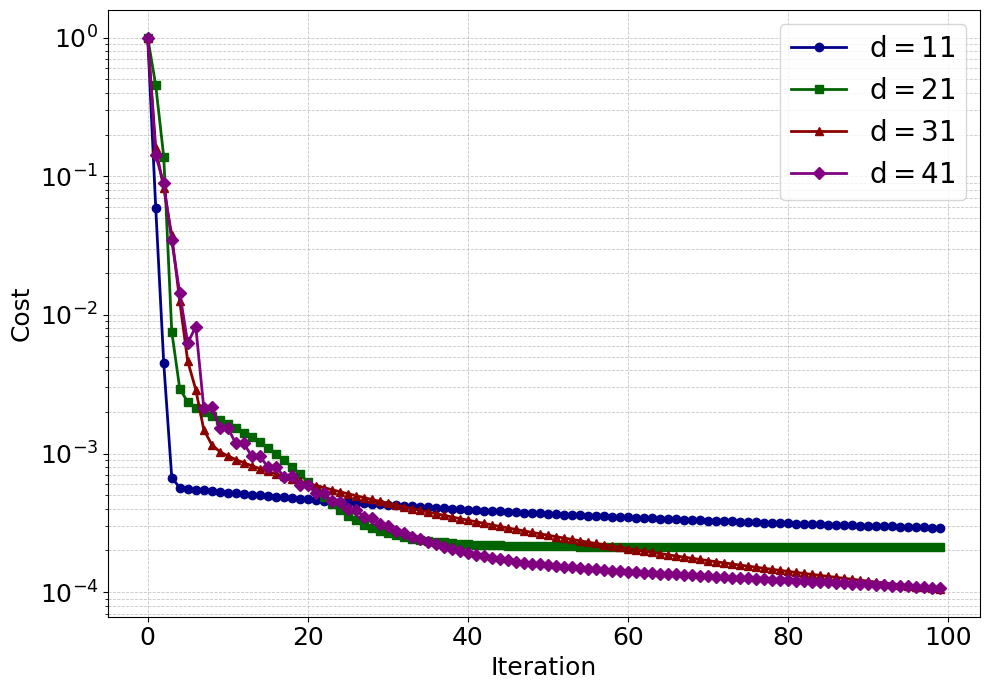}
    \caption{Logarithmic cost versus iteration for polynomial degrees 11, 21, 31, and 41 during 100 iterations. Phase angles are optimized using the Adagrad optimizer. Learning rate is 0.1. Each curve represents the cost value versus iterations for a specific polynomial degree in logarithmic scale.}
    \label{fig:optimization}
\end{figure}

Phase angles for QSVT were optimized using the Adagrad optimizer over 100 iterations to minimize the loss function defined in Eq. \ref{eqn:loss}. The loss function values at each iteration are presented in Fig. \ref{fig:optimization}. for polynomial degrees 11, 21, 31, and 41. Higher polynomial orders resulted in lower final cost values. 

\subsection{Polynomial approximation}
\begin{figure}[H]
    \centering
    \includegraphics[width=0.8\linewidth]{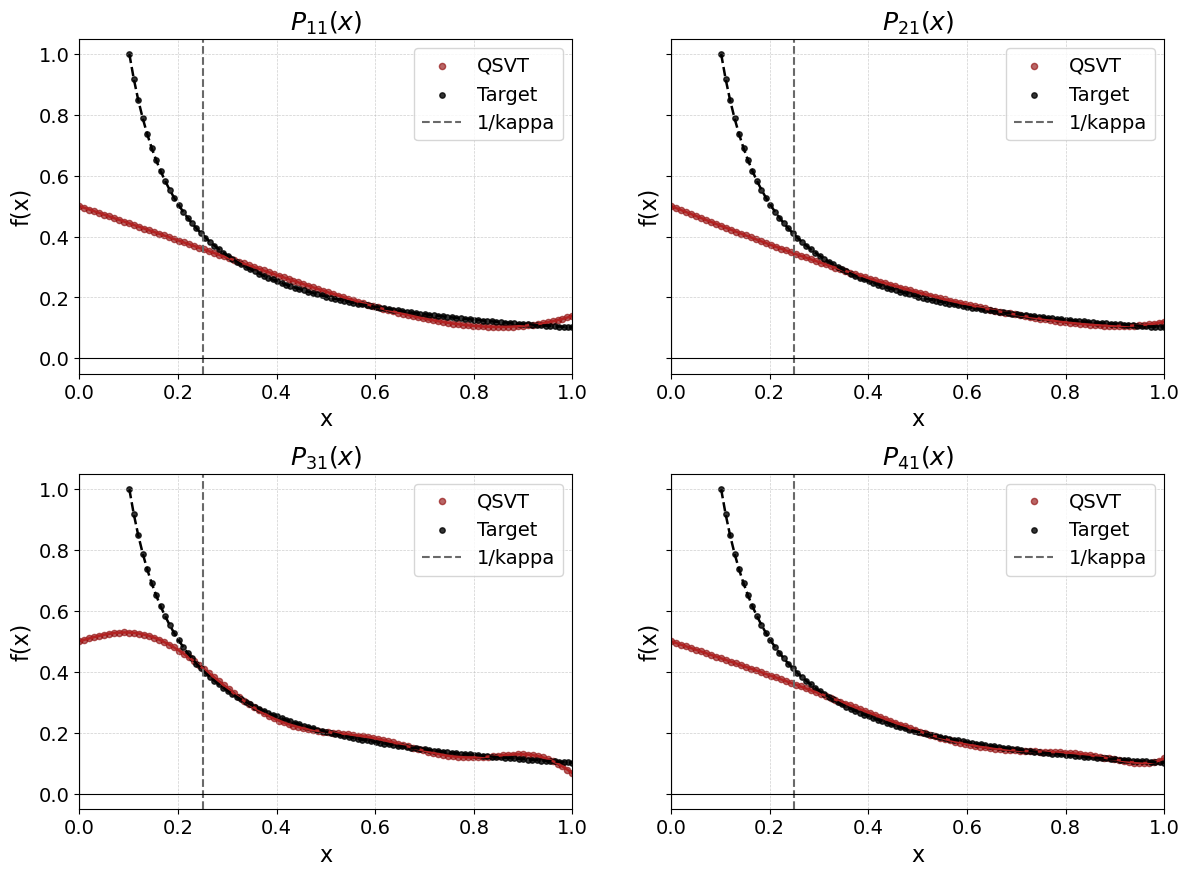}
    \caption{
        Comparison between the target inverse function and its QSVT approximation using polynomial degrees 11, 21, 31, and 41.
        Each subplot shows the QSVT output (red) and the target function (black), along with a vertical dashed line at $x = 1/\kappa$ where $\kappa = 4$.
        As the polynomial degree increases, the QSVT approximation improves and converges more closely to the target function.
    }
    \label{fig:targetVSqsvt}
\end{figure}

To evaluate the accuracy of the QSVT-based polynomial approximation of the inverse function, the \(\mathrm{L}_2\) relative error between the approximated and target functions was computed as
\begin{equation}
\label{eqn:l2_1}
\mathrm{L}_2 \,\text{Relative error} = \frac{\|f(x)_{\text{QSVT}} - f(x)_{\text{target}}\|_2}{\|f(x)_{\text{target}}\|_2}
\end{equation}
The errors decrease as the polynomial degree increases, demonstrating better alignment between the QSVT approximation and the target inverse function for higher polynomial orders. Table~\ref{tab:l2_errors} summarizes the \(\mathrm{L}_2\) relative errors as described in Eq. \ref{eqn:l2_1} for polynomial degrees 11, 21, 31, and 41. The comparison between the target inverse function and its QSVT polynomial approximations at different polynomial degrees is illustrated in Fig. \ref{fig:targetVSqsvt}.

\begin{table}[ht]
\centering
\caption{\(\mathrm{L}_2\) relative errors of the QSVT polynomial approximations for different polynomial degrees.}
\label{tab:l2_errors}
\begin{tabular}{c c}
\toprule
\textbf{Polynomial Degree} & \textbf{Relative \(\mathrm{L}_2\) Error} \\
\midrule
11 & 0.083404 \\
21 & 0.070187 \\
31 & 0.050324 \\
41 & 0.049560 \\
\bottomrule
\end{tabular}
\end{table}

\subsection{1D - Gaussian pulse in free space}
The case of a 1D Gaussian pulse propagating in free space, with periodic boundary conditions is being examined. The initial conditions for the electric and magnetic field are:

\begin{gather}
{E_{x}}_{0}(z,t=0) = \exp\left( -\frac{(z - 0.5)^2}{2 \cdot 0.05^2} \right) \label{eqn:Ez0} \\
{H_{y}}_{0}(z,t=0) = 0 \label{eqn:Hy0}
\end{gather}

\begin{figure}[H]
    \centering
    \includegraphics[width=1.0\linewidth]{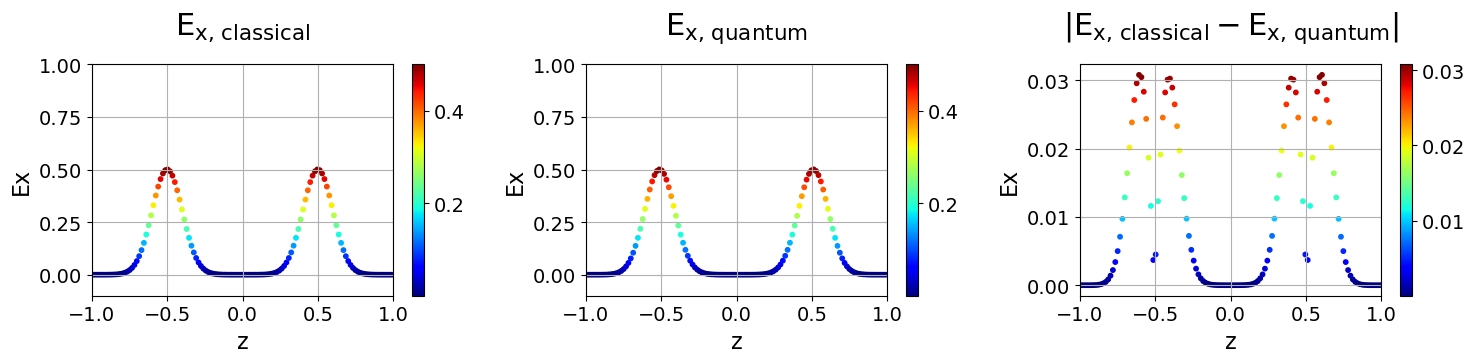}
    \caption{Electric field results for the 1D Gaussian pulse in free space for time t = 0.5, on 128 cells using 7 qubits. Classical scheme (left), QSVT solution (middle), and absolute error (right). A time step $\mathrm{\Delta t =0.01}$ was used for explicit Euler time stepping.}
    \label{fig:1D}
\end{figure}

The electric field results for 1D Gaussian pulse propagating in free space are shown for the time t = 0.5 in Fig \ref{fig:1D}. The initial electric and magnetic fields are defined as seen in Eq. \ref{eqn:Ez0} and Eq. \ref{eqn:Hy0}, respectively. The problem was solved on a grid of size $2^7 = 128$ cells, corresponding to 9 qubits as two of the qubits are ancillary qubits (from now on qubits refer to the system qubits unless stated otherwise). Classical scheme results are shown on the left, the QSVT results are shown at the middle, and the results for the absolute error between the classical scheme and the QSVT are shown on the right. A time step $\mathrm{\Delta t =0.01}$ was used for explicit Euler time stepping. According to Fig. \ref{fig:1D}, it can be understood that the error value is mostly affected by the $E_x$ amplitude, but only when the derivative doesn't approach 0.

To quantitatively assess the accuracy of the QSVT method compared to the classical scheme, we compute the \(\mathrm{L}_2\) relative error over time for the normalized solution, defined as
\begin{equation}
\label{eqn:l2_C-Q}
    \mathrm{L_2} \,\text{Relative error} = \frac{\| \mathbf{E}_\text{x, quantum} - \mathbf{E}_\text{x, classical} \|_2}{\| \mathbf{E}_\text{x, classical} \|_2}
\end{equation}
where \(\mathbf{E}_\text{x, quantum}\) and \(\mathbf{E}_\text{x, classical}\) are the electric field vectors obtained from the QSVT and classical schemes, respectively. 

\begin{figure}[H]
    \centering
    \includegraphics[width=0.5\linewidth]{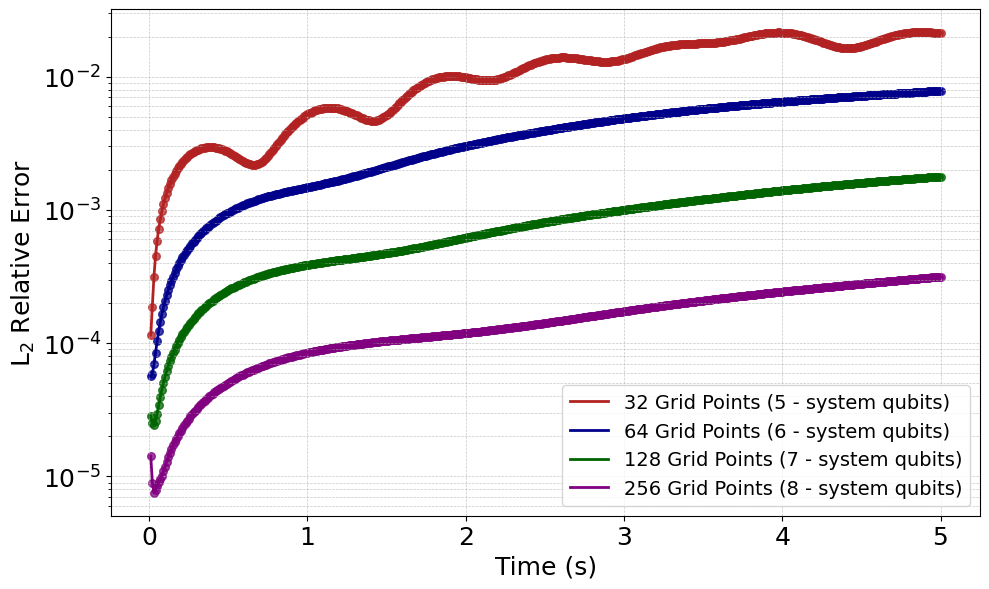}
    \caption{Relative $\mathrm{L}_2$ error the normalized solution over time between the classical scheme and QSVT for the 1D Gaussian pulse in free space. The results are shown for different grid sizes (32, 64, 128, and 256) corresponding to 5, 6, 7, and 8 qubits, respectively.}
    \label{fig:l2_error_qubits}
\end{figure}

Fig. \ref{fig:l2_error_qubits} presents the $\mathrm{L_2} \,\text{relative error}$ change in time for the normalized solution as described in Eq. \ref{eqn:l2_C-Q} for the 1D Gaussian pulse in free space across different spatial resolutions. The grid sizes correspond to 32, 64, 128, and 256 cells, which are simulated using 5, 6, 7, and 8 qubits respectively. Higher resolution yields lower relative error, indicating improved accuracy of the quantum scheme. However, the \(\mathrm{L}_2\) relative error increases  roughly linearly over time, which indicates a systematic error growth that prohibits long-term simulations.

\section{Discussion and Conclusion}
The case of a 1D Gaussian pulse propagating in free space, with periodic boundary conditions was examined. A QSVT-based quantum algorithm for solving $A\mathbf{x} = \mathbf{b}$ was simulated using PennyLane, demonstrating that a polynomial-based inverse function can be trained to produce accurate solutions with high overlap to the exact result.

The use of higher-degree polynomials yielded a closer approximation between the QSVT representation and the target inverse function. They also improved convergence during phase angle training. However, they led to slower convergence during phase angle training due to the increased complexity of the optimization landscape. The trade-off is increased circuit depth vs. accuracy. 

It was found that the $\mathrm{L_2}$ relative error between the classical and quantum solutions for the electric field grows linearly over time. This behavior is undesirable and suggests numerical instability or limitations in the QSVT-based method for long-time simulations. The accuracy of the solution is improved linearly in the number of qubits. However, this improvement comes at the cost of an increased circuit width.

It is important to note that current QSVT implementations rely on full knowledge of the quantum state, which limits their direct applicability on hardware beyond simulators. Even established tutorials on QSVT hardware implementation depend on access to the complete quantum state. Knowing the full state is problematic not only in Noisy Intermediate-Scale Quantum (NISQ) devices but also for fault-tolerant quantum computers, as complete state tomography requires $\mathcal{O}(2^{2n})$ measurements. A promising direction for future research is to identify applications or adaptations of QSVT that require only $\mathcal{O}(poly(n))$ observables, which could make the approach more practical, especially for near-term quantum devices.

This study demonstrates how a classical compact scheme can be integrated with quantum solvers such as QSVT. Other schemes and methods such as the Crank-Nicolson scheme, compact filtering schemes, high-order compact schemes, and high-order upwind summation-by-parts methods\cite{Liu2022, VISBAL2002155, LELE199216, Ranocha_2025} can be integrated with QSVT easily using the same methods.

\bibliographystyle{unsrt}
\bibliography{references}

\end{document}